%% file: main.tex
\documentclass[journal]{vgtc}                

\usepackage{mathptmx}
\usepackage{graphicx}
\usepackage{times}
\usepackage{comment}
\usepackage{enumitem}
\usepackage{amsmath}
\usepackage[normalem]{ulem}
\usepackage{xcolor}
\usepackage[bookmarks,backref=true,linkcolor=black]{hyperref} 
\hypersetup{
  pdfauthor = {},
  pdftitle = {},
  pdfsubject = {},
  pdfkeywords = {},
  colorlinks=true,
  linkcolor= black,
  citecolor= black,
  pageanchor=true,
  urlcolor = black,
  plainpages = false,
  linktocpage
}

\onlineid{0}

\vgtccategory{Research}

\vgtcinsertpkg

\title{Bayesian-Assisted Inference from Visualized Data}

\author{Yea-Seul Kim, Paula Kayongo, Madeleine Grunde-McLaughlin and Jessica Hullman}
\authorfooter{
\item
 Yea-Seul Kim is with the University of Washington. E-mail: yeaseul1@uw.edu.
\item
  Paula Kayongo is with Northwestern University. E-mail: paulakayongo2023@u.northwestern.edu
  
\item 
    Madeleine Grunde-McLaughlin is with the University of Pennsylvania. E-mail: madeleine.grunde@gmail.com
\item
  Jessica Hullman is with Northwestern University. E-mail: jhullman@northwestern.edu
  
}
\shortauthortitle{Biv \MakeLowercase{\textit{et al.}}: Global Illumination for Fun and Profit}

\abstract{    \input{0_abstract.tex}

} 


\CCScatlist{ 
 \CCScat{K.6.1}{Management of Computing and Information Systems}%
{Project and People Management}{Life Cycle};
 \CCScat{K.7.m}{The Computing Profession}{Miscellaneous}{Ethics}
}

  \teaser{
 \centering
 \includegraphics[width=16cm]{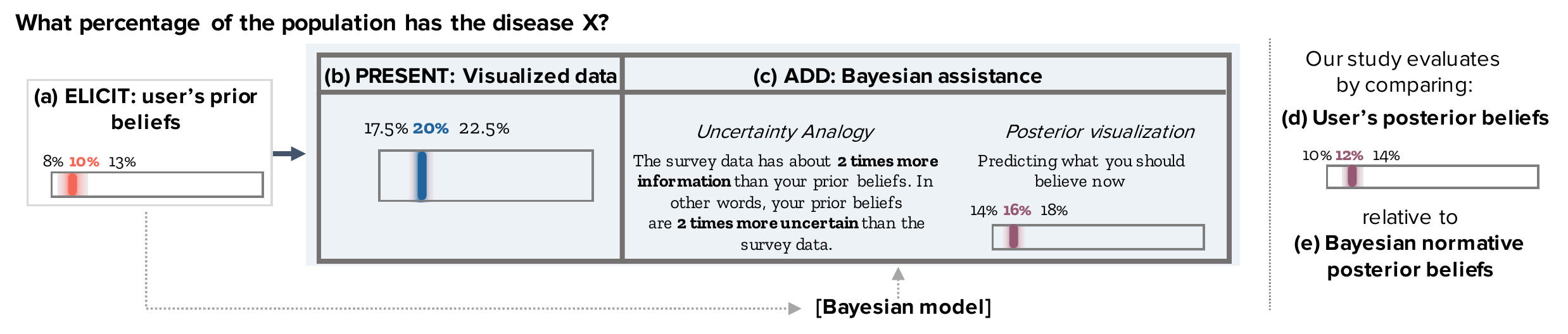}
 \vspace{-3mm}
  \caption{
  Using Bayesian inference to assist how data is shown to improve belief updating. (a): The viewer holds prior beliefs about a parameter such as a disease rate in the population, which are elicited in the form of a probability distribution. (b): The user is presented with an observed dataset estimating the rate, which conveys information about the likelihood function. (c): The observed data is accompanied by Bayesian assistance techniques in the form of an uncertainty analogy or visualization of Bayesian posterior predictions derived from their prior beliefs and a normative Bayesian model. (d): In our experiment, we elicit posterior beliefs and use the deviation between these beliefs and (e) the normative beliefs to evaluate the two types of Bayesian assistance. The goal of Bayesian assistance is to bring the user's updated beliefs closer to (e). 
  }
\label{fig:bayesian}
  }

\definecolor{bias}{HTML}{ff6600}
\newcommand{\bias}[1]{\textbf{\textcolor{bias}{#1}}}

\definecolor{disp}{HTML}{51A5BA}
\newcommand{\disp}[1]{\textbf{\textcolor{disp}{#1}}}

\newcommand{\jessica}[1]{\textbf{\textcolor{red}{JH:#1}}}
\newcommand{\yeaseul}[1]{\textbf{\textcolor{cyan}{#1}}}

\begin{document}



\maketitle
\input{1_intro.tex}
\input{2_relatedWork.tex}
\input{3_motivation.tex}
\input{4_study.tex}

\input{5_results.tex}
\input{6_discussion.tex}

\bibliographystyle{abbrv}
\bibliography{template}
\end{document}

%% file: 0_abstract.tex
A Bayesian view of data interpretation suggests that a visualization user should update their existing beliefs about a parameter's value in accordance with the amount of information about the parameter value captured by the new observations. Extending recent work applying Bayesian models to understand and evaluate belief updating from visualizations, we show how the predictions of Bayesian inference can be used to guide more rational belief updating.
We design a Bayesian inference-assisted \textit{uncertainty analogy} that numerically relates uncertainty in observed data to the user's subjective uncertainty, and a \textit{posterior visualization} that prescribes how a user should update their beliefs given their prior beliefs and the observed data.
In a pre-registered experiment on 4,800 people, we find that when a newly observed data sample is relatively small (N=158), both techniques reliably improve people's Bayesian updating on average compared to the current best practice of visualizing uncertainty in the observed data. 
For large data samples (N=5208), where people's updated beliefs tend to deviate more strongly from the prescriptions of a Bayesian model, we find evidence that the effectiveness of the two forms of Bayesian assistance may depend on people's proclivity toward trusting the source of the data. We discuss how our results provide insight into individual processes of belief updating and \textit{subjective} uncertainty, and how understanding these aspects of interpretation paves the way for more sophisticated interactive visualizations for analysis and communication.

%% file: 1_intro.tex
\section{Introduction}
People look to data visualizations in the media, government, and science to help them form beliefs about the world around them. However, abundant research indicates that people often struggle to properly account for uncertainty in making judgments from data. For example, many people overinterpret small samples~\cite{monmouth,monmouth2}. 
In other cases they may underreact to data, misjudging how informative large samples are~\cite{ambuehl2018} or failing to update their beliefs when a sample conflicts with their pre-existing beliefs~\cite{callaghan}. 

Cognitive errors like under- and overreaction to data can be defined by comparing human judgments to Bayesian inference, a statistical method that prescribes how to update probabilistic beliefs given new evidence. Imagine you are interested in a political candidate A's chance of winning an election, and you have some expectations about that chance, based on, for example, seeing early results from a small poll of registered voters, and your experiences talking to others in your social circle. If asked to describe your beliefs, you'd say your best guess of the candidate's chance of winning the election is 51\%, with a 95\% chance that the value will be between 47\% and 55\%. In a Bayesian framework, these beliefs are called your \textit{prior beliefs}.

One day you encounter a visualization of new poll results. The data indicates that A has a 60\% chance of winning, based on responses from around 1000 people, with the chance of winning falling between 57\% and 63\% with high confidence (e.g., 95\%). What should you believe after encountering the second poll? The laws of Bayesian belief updating prescribe an ``optimal'' way for combining prior and new information. Assuming that you have no reason to distrust the new evidence, you should update your beliefs proportional to the amount of new information that the poll provides over what you already believed. Bayesian inference formalizes this intuition through Bayes rule, which states that your \textit{posterior beliefs} about a parameter after observing new data are proportional to your prior beliefs about the parameter multiplied by the information contained in the new evidence about the parameter. In this case, your new beliefs about A's chance of winning should be around 57\%, with a 95\% interval between 54\% and 59\%.

Recent work shows how a Bayesian cognition perspective can deepen understanding of visualization interpretation~\cite{kim2019,wu2017} and contribute to more rigorous evaluation, in which deviation from Bayesian updating is used as a proxy for understanding which visualizations best support accurate perception of how informative data is~\cite{kim2019}. We extend this work by considering the generative potential of predictions from models of Bayesian inference to \textit{guide} belief updating from visualized data.
We propose two \textit{Bayesian assistance} techniques that use the mathematical intuitions of Bayesian theory to guide a user's belief formation process as they interact with visualized data. Both techniques treat the user's subjective uncertainty about a parameter value before seeing newly observed data (i.e., their prior distribution) as a reference point against which the uncertainty in the observed data can be compared (Fig.~\ref{fig:bayesian}b2). An \textit{uncertainty analogy} relates uncertainty in observed data to uncertainty in the user's prior. A \textit{posterior visualization} depicts the posterior beliefs predicted by Bayesian inference given the user's prior beliefs.

How does Bayesian assistance change users' beliefs as they interact with a visualization? We present a preregistered experiment with 4,800 participants in which we compare users' belief updating under Bayesian uncertainty analogies and posterior visualizations to beliefs based on common presentations of uncertain estimates like point estimates with reported sample size or a shaded interval displaying probability density. We find that:
\begin{itemize}[noitemsep,nolistsep,leftmargin=*]
    \item For small datasets (N=158), both techniques bring the average user's belief updating closer to normative Bayesian inference.
    \item Eliciting a prior from a user itself can encourage more Bayesian updating, as evidenced through an aggregate analysis of people's updating without and without elicitation.  
\end{itemize}

We conclude by discussing the implications of our results as well as the adoption of Bayesian inference to guide visualization design and evaluation.

%% file: 2_relatedWork.tex
\section{Related Work}
\subsection{Visually Communicating Uncertainty}
\label{sec:related}
Research in judgment and decision-making demonstrates how human judgments under uncertainty can diverge from statistical accounts. For example, belief in the law of small numbers describes how many people are too confident in the representativeness of small samples~\cite{tversky1971}. More recent work describes how a related bias called non-belief in the law of large numbers, in which a person simply believes that proportions in \textit{any} given sample might be determined by a rate different than the true rate (i.e., misunderstands the relation between sample size and error), is compatible with the earlier work on small samples by Tverksy, Kahneman, and many others~\cite{benjamin2016}. 

Some interventions can reduce biases in interpreting uncertainty. Research in uncertainty visualization has proposed many techniques for visually representing quantified uncertainty distribution to improve judgments or decisions, from boxplots (e.g.,~\cite{potter2012}) to visualizations of probability density as area, shading or other visual properties (e.g.,~\cite{correll2014,feng2010}) to frequency-based representations of probability like hypothetical outcome plots and quantile dotplots ~\cite{bastin2002,ehlschlaeger1997,fernandes2018,hofman2020,hullman2015,kale2018,kay2016ish} probabilistic animations. We compare how well users update their beliefs using two Bayesian assistance relative to a conventional interval and shaded density representation of a dataset.



\subsection{Bayesian Inference in Judgments \& Decisions}
Empirical research in economics and mathematical cognition demonstrates the role of beliefs in numerous judgments and decisions. Manski~\cite{manski2018} argues against a long standing bias in economics toward inferring beliefs from choice, noting that eliciting probabilistic beliefs provides useful and predictive insight into behavior. He surveys economic literature on how people form beliefs and how these beliefs influence their financial decision making~\cite{bailey2018,armona2017} or other consumption~\cite{bachmann2015, d2016, roth2018,armantier2017,binder2018,cavallo2017,coibion2018,fuster2012, ferrell1980,zimmer1983}.
Camerer~\cite{camerer1995individual}, Schotter and Trevino~\cite{schotter2014} summarize the value of studying beliefs from laboratory findings, while Abeler et al.~\cite{abeler2019} use quantitative meta-analysis to show that experiment subjects can generally be trusted to report honest beliefs in economics experiments~\cite{abeler2019}.

Mathematical psychologists have shown how Bayesian models of cognition help explain a range of perceptual and cognitive phenomena, such as inferring causal relationships~\cite{steyvers2003,sobel2004} or inductive learning~\cite{tenenbaum2006,griffiths2006}. For example, Griffiths and Tenenbaum~\cite{griffiths2006} demonstrate that the aggregate posterior belief distribution across people approximates the normative Bayesian posterior over various ``everyday quantities'' such as cake baking times and human lifespans. 

Though the authors explicitly suggest that a mathematical account would not be feasible, McCurdy et al.'s~\cite{mccurdy2018} suggestion that implicit error captures how users ``mentally adjust'' data-driven estimates in interpretation resembles the Bayesian ideal that prior beliefs influence inferences drawn from new data. In contrast to their assertion that mathematical frameworks are not possible, we demonstrate how Bayesian modeling can combine subjective beliefs with observed data to reduce integration errors that may arse in mental approximation.

Until recently, research on the role of visualizations in promoting Bayesian reasoning was limited to studying how visualizations affect performance on classic conditional probability tasks like the mammography problem~\cite{micallef2012,ottley2015,garcia2013,tsai2011,ottley2012,gigerenzer1995,cole1989}. However, several recent visualization studies apply Bayesian modeling to visualization interpretation~\cite{kim2019,wu2017}. In the closest prior work, Kim et al.~\cite{kim2019} presented people with survey estimates of several proportions, finding that at an individual level, people's posterior beliefs diverged considerably from normative Bayesian. In aggregate, however, people's posterior beliefs closely approximated the predictions of normative Bayesian inference for estimates based on small samples (N=158), but not for those based on very large samples (N=750k). Kim et al. show how the deviation between a person's posterior beliefs and the Bayesian normative posterior beliefs can be used as a proxy for a user's uncertainty comprehension. Our work extends this inquiry by considering whether Bayesian inference can also be used to \textit{generate} personalized data presentations based on a user's prior beliefs.

%% file: 3_motivation.tex
\section{Motivating Bayesian Assistance}

We introduce the assumptions behind applying a Bayesian perspective to visualization interpretation, then the specific components of our Bayesian modeling approach in the context of a belief updating scenario.

\subsection{Assumptions of a Bayesian Approach to Visualization}
\label{sec:assume}
To apply Bayesian inference to visualization, we assume that prior to interacting with a visualization, a user has some state of prior beliefs about a parameter which the data provides an estimate of (e.g., a rate). We assume that any user's prior beliefs can be elicited through an interactive interface, and represented by a probability distribution. We can think of how tightly concentrated this distribution is as the strength of the user's beliefs, capturing how confident they are in their knowledge about the parameter value. The user's prior beliefs about a parameter can range from no relevant knowledge about the parameter value (e.g., a uniform distribution in which all values of the parameter are thought to be equally likely) to near complete certainty (e.g., high confidence that the value is within a very small interval).

We assume that the user will update their prior beliefs about the parameter upon viewing new information in a visualization. We assume that the closer the user's belief update is to optimally combining the information in their prior with the new visualized data (as defined by a standard Bayesian model of updating a sample proportion), the more rationally they have updated their beliefs.

For example, if one has no reason to believe that any particular value of the parameter is more likely than any other, their posterior beliefs should be equal to the evidence that the visualized data provides about the parameter value. If they had very strong prior beliefs about the parameter, and saw a relatively small amount of evidence in the visualization, their posterior beliefs should remain close to even identical to their prior beliefs.

To model this process we use mathematical formulations standard in Bayesian statistics, including to fit the elicited beliefs to a statistical (prior) distribution, to represent the information about the parameter implied by the dataset (likelihood), and to calculate the Bayesian posterior beliefs. We provide further mathematical details below.

Finally, note that Bayesian inference in cognition is typically assumed to be the implicit process; our work explores whether making predictions from normative Bayesian updating \textit{explicit} can be beneficial to users. Further, unless possible bias is intentionally modeled, a Bayesian model of updating will assume that prior beliefs and observed data are equally credible sources of information. Our work demonstrates how people's self-reported trust in data's credibility helps predict where this assumption may not hold.

\subsection{Applying Bayesian Inference to Visualization Scenario}
\label{sec:applying}
Consider a scenario in which a user will be presented with a visualized estimate of a parameter $\theta$. Imagine that the parameter is the proportion of residents of U.S. assisted living centers who have Alzheimer's. As a proportion, $\theta$ can theoretically take any value from 0 to 1. Before the user views \textit{observed data}, they articulate their prior beliefs by assigning probability over plausible values of $\theta$ using an interactive interface (Fig~\ref{fig:bayesian}a).

In Bayesian inference, beliefs take the form of a probability distribution. For a proportion parameter $\theta$, a Beta distribution is a convenient distribution to capture beliefs. Two parameters sufficiently define a unique Beta distribution: $Beta(\alpha, \beta)$. We can think of $\alpha - 1$ as the number of successful events (e.g., the number of residents in assisted living centers who are believed to have Alzheimer's), and $\beta - 1$ as the number of unsuccessful events (e.g., the number of residents in assisted living centers who are believed to not have Alzheimer's).

Imagine a user who guesses that approximately 10\% of residents in assisted living centers have Alzheimer's, but with relatively high uncertainty. Assume that the information their beliefs imply is equivalent to having observed a sample of 10 assisted living center residents, one of which had dementia. Their prior beliefs are captured by the distribution $Beta(2, 10)$. The sum of the successful events and the failure events (i.e., 10) represents the amount of information (or conversely uncertainty) contained in the user's prior distribution.

Imagine that the user is next presented with a visualization of an estimate captured by observed data (Fig~\ref{fig:bayesian} (b1)), such as the proportion of assisted living center residents with dementia according to records for a chain of centers with locations across the country. Out of 1,000 residents of these chains, 420 have dementia. We model the data generating process as a binomial process in which any individual independently has the disease with a certain (identical) probability $\theta$. 

We represent the observed data as a \textit{likelihood function} capturing the probability of different values of $\theta$ given the observed data. Conveying a sense of likelihood is the goal of most approaches to communicating uncertainty in estimates. The likelihood encodes the relative number of ways that different values of $\theta$ could produce the observed proportion given our assumptions about the data generating process and the size of the observed sample. The likelihood function for a sample proportion, 42\%, of 1,000 total residents can be represented by $Binomial(1000, 0.42)$, implying an expected 420 successful events and 580 failure events but with some uncertainty due to sampling error.

\vspace{-4mm}
\begin{equation}
\begin{split} 
\# of successes_{posterior} = \# of successes_{prior} + \# of successes_{data} \\ \# of failures_{posterior} = \# of failures_{prior} + \# of failures_{data} 
\end{split}
\label{eq:simple_bayesian}
\end{equation}

The \textit{normative posterior distribution} (Fig~\ref{fig:bayesian}e) that predicts rational updating is calculated by using Bayes rule to update the probability of $\theta$ in the prior with the information about $\theta$ implied by the likelihood function. 
Equation~\ref{eq:simple_bayesian} results from using Bayes rule to estimate the number of successful events and the failure events in the posterior beliefs as a function of the estimates implied by the observed data and prior. The number of successful and failure events in the posterior beliefs is equivalent to a Beta distribution: $Beta(422, 590)$. Intuitively, under Bayesian inference the user's belief distribution after encountering the observed data shifts proportionally to the amount of information contained in the two distributions.

\subsection{Designing Bayesian Assistance}
We propose two \textit{Bayesian assistance techniques} that exploit the user's prior beliefs. An \textit{uncertainty analogy} relates uncertainty in observed data to uncertainty in the user's prior, and a \textit{posterior visualization} depicts the posterior beliefs predicted by Bayesian inference, given the user's prior beliefs. 

\subsubsection{Uncertainty Analogy}
The user's prior distribution captures their uncertainty about the parameter value before seeing the observed data. We can treat this subjective uncertainty as a personally meaningful reference against which uncertainty in the observed data can be compared.
Imagine you are presented with a visualization and text telling you how much information the visualized data contains relative to how informed you were about the topic already: ``Your prior beliefs have 2 times more information than the data.'' 

To generate the multiplicative factor, we compare $\kappa$ (a proxy for sample size defined as $\alpha$ + $\beta$) in the prior distribution ($\kappa_{prior}$) to the sample size of the observed data ($\kappa_{data}$). To avoid multipliers less than one, we always chose the distribution (Beta corresponding to likelihood or participant's prior) for which $\kappa$ was lower as the reference distribution. For example, if $\kappa_{data}$ is greater than $\kappa_{prior}$, we calculated the multiplier as $\kappa_{data}/\kappa_{prior}$ (e.g., \textit{Your prior beliefs have 2 times more information than the data}), calculating the multiplier as $\kappa_{prior}/\kappa_{data}$ in the case where $\kappa_{prior}$ was greater.

\subsubsection{Posterior Visualization}
An even more direct way to guide a user toward Bayesian inference is to present them with the normative belief distribution calculated using their prior beliefs and the likelihood. Imagine that in addition to an observed dataset, you are presented with a visualization suggesting how you should update your beliefs, in the form of the normative posterior calculated using your prior distribution, along with a brief explanation of how it was derived (i.e., by combining the information in their prior beliefs with that in the observed data).

%% file: 4_study.tex
\section{Experiment: Bayesian Assistance}
\label{sec:experiment}
We designed and preregistered a large crowd-sourced between-subjects experiment to evaluates how participants' appear to update their beliefs under Bayesian assistance versus more conventional depictions of proportion estimates. 

\begin{figure}[htb]
 \centering
  \includegraphics[width=\columnwidth]{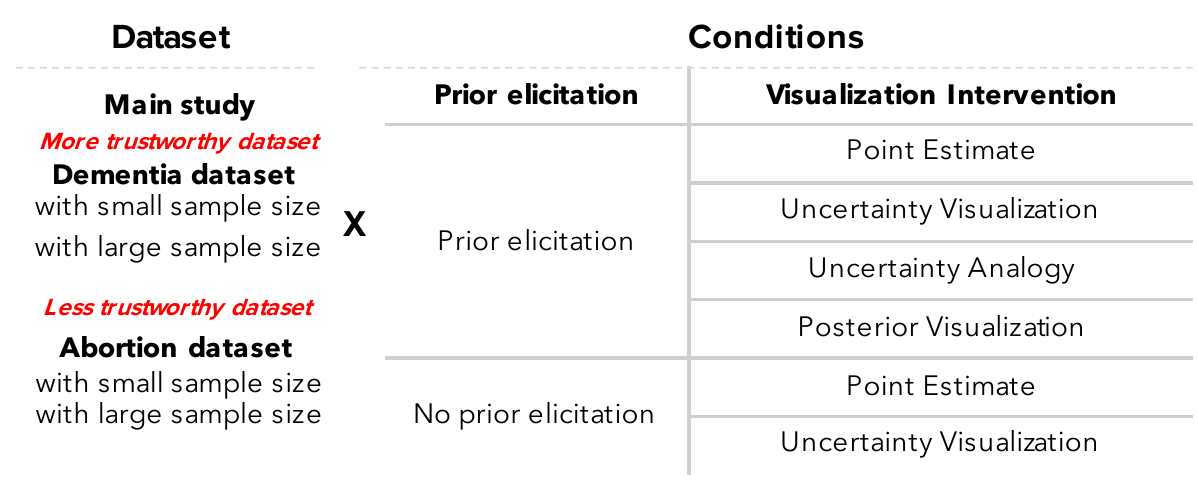}
  \vspace{-6mm}
  \caption{The study conditions and datasets. }
    \vspace{-0.1in}
  \label{fig:conditions}
\end{figure}

\subsection{Study Conditions \& Research Questions}
We tested four approaches to conveying uncertainty (Fig.~\ref{fig:conditions}).

\renewcommand\labelitemi{\tiny$\bullet$}
\begin{itemize}[nolistsep,leftmargin=*]
    \setlength\itemsep{0.2em}
    
    \item \textbf{Point Estimate (with sample size)}: Participants view a point estimate of the observed proportion with the size of the sample in text. 
    
    \item \textbf{Uncertainty Visualization}: Participants view a point estimate of the observed proportion along with a probability density shaded interval in which the estimate is expected to fall with high probability (95\%). 
    
    \item \textbf{Uncertainty Analogy}: Participants view the uncertainty visualization alongside an uncertainty analogy. A brief explanation of how the analogy was generated (e.g., ``We directly compared the sample size of the study to the sample size implied by your prior beliefs.'') is also presented.
    
    \item \textbf{Visualization}: Participants view the uncertainty visualization alongside a visualization of the normative posterior distribution. A brief explanation of how the posterior was arrived at (including an analogy expression comparing the uncertainty in the participant's prior beliefs to that of the data as above) is presented.

\end{itemize}
\begin{figure}[h!]
 \centering
  \includegraphics[width=\columnwidth]{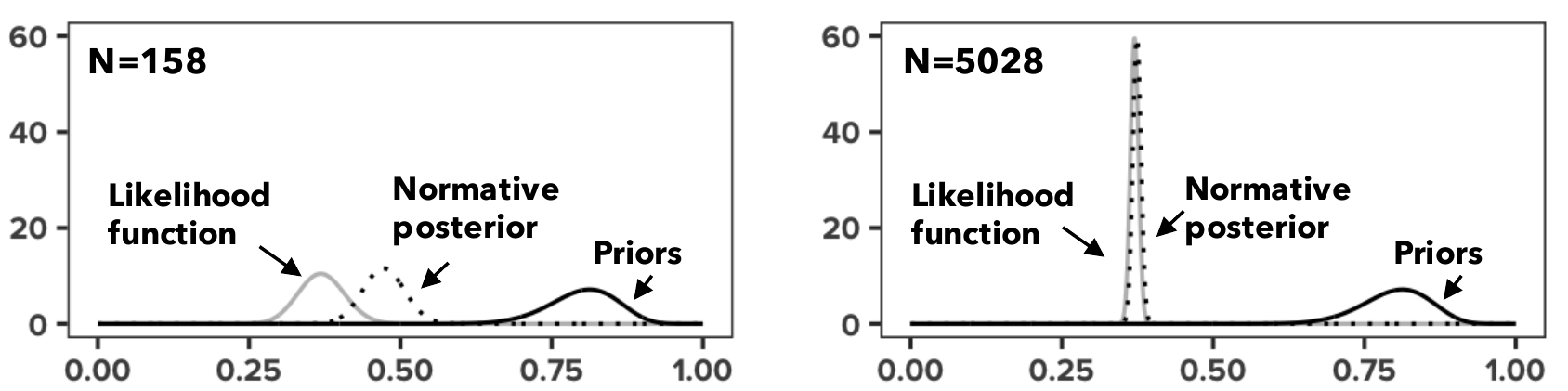}
  \vspace{-6mm}
  \caption{Illustration of how normative posterior beliefs (dashed) are influenced by the sample size of the observed data (represented by the likelihood in gray) given a prior distribution (solid). Assuming a relatively weak prior, when the sample size is small, the normative posterior distribution is located between the likelihood and the prior. Assuming the same prior and a large sample observed dataset, the normative posterior distribution is nearly identical to the likelihood function.}
  \vspace{-0.15in}
  \label{fig:sample_size_simulation}
\end{figure}

\subsubsection{Robustness to Varying Sample Size}
As Fig.~\ref{fig:sample_size_simulation} left shows, a weak prior belief distribution still has a demonstrable impact on the normative posterior beliefs when the observed data is relatively small (N=158). For a larger sample (N=5208) the normative posterior distribution is nearly identical to the observed data (Fig.~\ref{fig:sample_size_simulation} right).
By varying sample size, we use our experiment to investigate whether a tendency for people's posterior beliefs to deviate more substantially from the normative posterior distribution for large samples found in prior work~\cite{kim2019} holds for our participants as well. We chose 158 (after Kim et al~\cite{kim2019}) and 5,208 as samples in the low thousands are common in presentations of poll or survey results that people encounter in everyday life.

\subsubsection{Robustness to Topic Controversy}
Besides misunderstanding uncertainty, not trusting that a dataset is a faithful depiction of reality is another possible reason for the deviation between one's posterior beliefs and the normative Bayesian posterior.

To investigate the impact of the perceived ``controversialness'' of data on the effects of Bayesian assistance, we identified two datasets that vary in how likely they are to be perceived as having been manipulated. We recruited 200 Mechanical Turk workers in the U.S. with approval ratings of 97\% and above. Participants viewed pairwise combinations of six datasets: the proportion of 1) residents of U.S. assisted living centers residents who have Alzheimer's or other dementia, 2) corn production relative to other grain production in the U.S., 3) patients in the U.S who misuse opioids prescribed for chronic pain, 4) foreign-born residents in the U.S., 5) adults in the U.S who think third trimester abortion should be illegal regardless of circumstances, and 6) adults in the U.S. who support the death penalty.

In a first session, on each trial the participant saw a pair of dataset descriptions (i.e., a summary of the variable) side by side. Participants were asked to choose one dataset that ``seems more likely to be tampered with or manipulated to persuade'' using a radio button. Participants viewed a total of 15 pairs (trials). In the second session, participants viewed the same 15 pairs but where the original proportion from the source is presented with a 95\% highest density interval calculated by for an assumed sample size of 158. We randomized the order of pairs in both sessions.

We ranked the datasets by perceived manipulation using the sum of participants' votes per dataset. The proportion of U.S. assisted living centers residents who have Alzheimer's obtained the fewest votes across both questions, while the proportion of Americans who believe long-term abortions should be illegal unilaterally obtained the most.

\subsubsection{Impact of Prior Elicitation}
It is possible that prior elicitation itself may affect how ``Bayesian'' a person appears to be, for example if it encourages the user to be more sensitive to uncertainty in the data. 
We include two conditions for which we do \textit{not} elicit prior beliefs--No Elicitation-Point Estimate and No Elicitation-Uncertainty Visualization--and use them to evaluate the impact of elicitation on deviation from normative Bayesian belief updating. Though individual-level updating with and without elicitation cannot be directly compared without eliciting the individual's prior, an aggregate-level analysis, in which we assign No Elicitation conditions a common prior learned from many participants, allows us to observe how elicitation appears to change updating at an aggregate level.

\subsection{Experiment Design \& Procedure}
We ran our experiment as a between-subjects study. Participants were randomly assigned to one of the six elicitation and visualization conditions and one of four datasets (small or large dementia dataset or small or large abortion data) (Fig.~\ref{fig:conditions}). We pre-registered our conditions, sample sizes, and analysis\footnote{\href{https://aspredicted.org/blind.php?x=sq3xz8}{Pre-registration I}, \href{https://aspredicted.org/blind.php?x=2uc84m}{Pre-registration II}}. An introductory page described the dementia datasets (originally from the U.S. National Center for Health Statistics~\cite{nytElderly}) as having been collected by a national health agency, and the abortion datasets (originally from FOX News~\cite{foxnews}) as having been collected by a media outlet. 

 \begin{figure}[h!]
 \centering
  \includegraphics[width=\columnwidth]{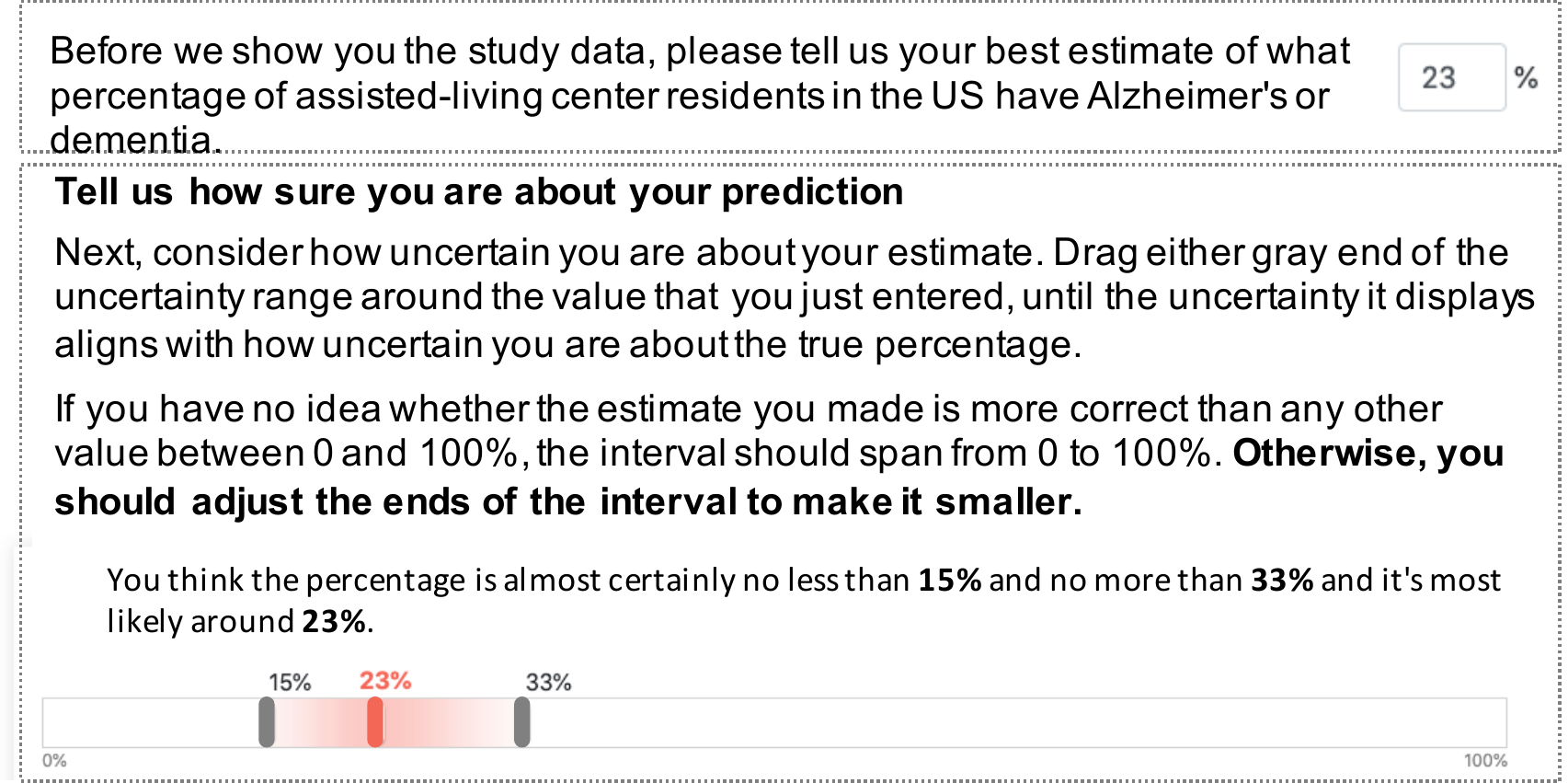}
  \vspace{-5mm}
  \caption{The elicitation interface. First, the participant enters a point estimate (top), then they specify how certain they are about their estimate by dragging either end of the interval (bottom). When the participant interacts with either handle, the other handle updates to accommodate the updated Beta distribution.} 
  \label{fig:interface_elicitation}
\end{figure}

\begin{figure*}[t]
 \centering
  \includegraphics[width=0.93\linewidth]{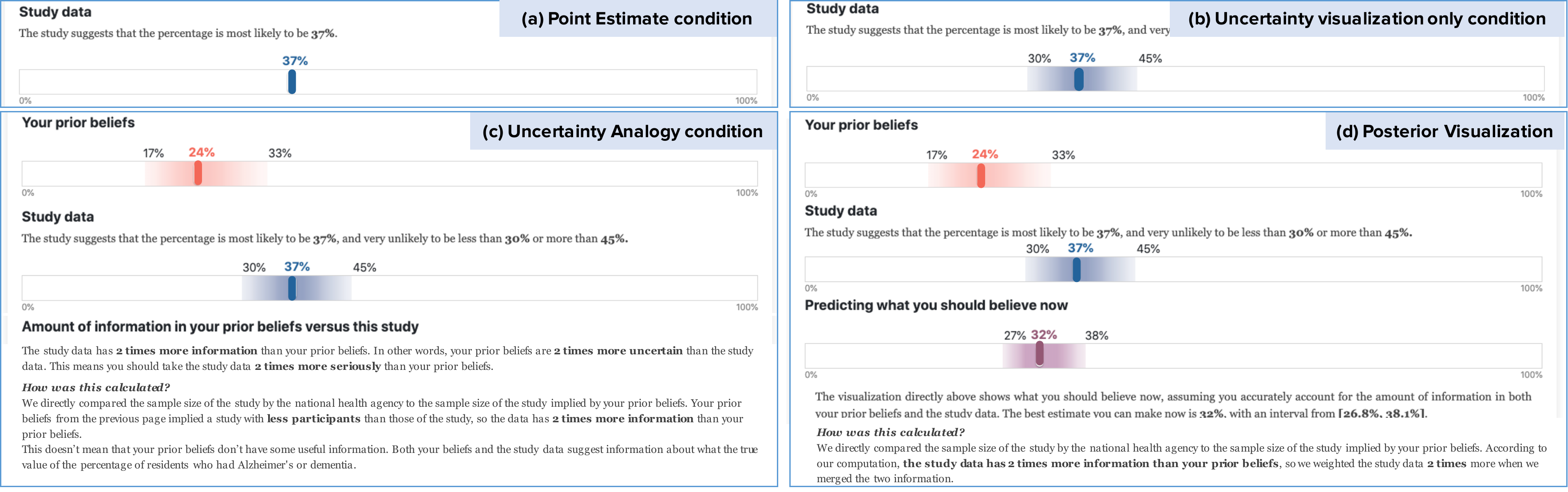}
  \vspace{-4mm}
  \caption{Conditions in our experiment, including visualizing observed data as a point estimate with sample size, using a high probability interval with shading to visualize uncertainty in the observed data only, providing an uncertainty analogy based on the participant's prior, and providing a predicted posterior visualization based on the user's prior.
  }
  \vspace{-3mm}
  \label{fig:stimuli}
\end{figure*}

\subsubsection{Prior Belief Elicitation}
Participants assigned to elicitation conditions first provided their prior beliefs (Fig.~\ref{fig:interface_elicitation} top). 
We designed an interface that prompted the participant to enter their best estimate of the parameter of interest (e.g., the percentage of assisted-living center residents in the US have Alzheimer's or dementia), following prior research in proportion prior elicitation from experts~\cite{wu2008}. 
A two-handled slider then appeared, representing an interval around the value they provided as their estimate, with endpoints at 0 and 100\%. Participants were asked to specify a range around the value by dragging the end of the interval until its width aligned with how uncertain they felt about the true rate (Fig.~\ref{fig:interface_elicitation} bottom). Participants were explicitly told that if their estimate represented a truly random guess, then their interval should span from 0 to 100\%; otherwise they should adjust the ends of the interval to make it smaller. When the participant interacted with either handle, we updated the concentration parameter ($\kappa$) based on the handle's value and the mode, then calculated the other handle's location to reflect the 95\% interval of the new Beta distribution. Specifically, $\kappa$ is inversely proportional to the width of the elicited interval. Text above the slider reflected the specified prior (e.g., \textit{You think the percentage is almost certainly no less than} \textit{15\%} \textit{and no more than} \textit{33\%} \textit{and it's most likely around} \textit{23\%}, Fig.~\ref{fig:interface_elicitation}c).

\subsubsection{Presentation of Observed Data}
After prior elicitation, all participants examined the observed data. To create the visualization stimuli, we used the proportions from the original source of the datasets (dementia dataset: 42\%, abortion dataset: 37\%) and varied the sample size that a participant was assigned (small: 158, large:5208).
Participants in the Point Estimate conditions saw the point estimate of the proportion plotted with the number of successes and sample size in text only (Fig~\ref{fig:stimuli}a). Participants in the Uncertainty Visualization and Bayesian assistance conditions saw the point estimate plotted with an interval depicting the lower and upper bound of the corresponding Beta distribution for the Binomial likelihood function, with shading proportional to probability density (Fig~\ref{fig:stimuli}b).

\subsubsection{Presentation of Bayesian Assistance}
After viewing the data and prior visualization, participants in the assistance conditions then clicked for the Bayesian assistance, which appeared below the visualization of the observed data. For participants in the Analogy condition, we presented an analogy in text (Fig.~\ref{fig:stimuli}c). For participants in the Posterior Visualization condition, we presented a visualization like our uncertainty visualization of the observed data, but where the distribution shown is the Beta distribution corresponding to the predicted posterior from our Bayesian model (Fig.~\ref{fig:stimuli}d).

\subsubsection{Posterior Belief Elicitation \& Post-Task Questions}
All participants then submitted their posterior beliefs on the next screen. On a final screen, participants were asked demographic questions (gender, education level, and age), and how likely they thought it was that the data was manipulated on a five-point Likert scale with endpoints labeled Not at all likely (1) and Extremely likely (5). 
The final screen asked participants what proportion corresponded to the observed data they had been shown via multiple choice (Below 30\%, between 30\% to 60\%, above 60\%) as a preregistered exclusion criteria to filter participants who were not paying attention from analysis.

\subsubsection{Participants}
We recruited participants on Amazon Mechanical Turk, removing those who failed the preregistered exclusion criteria question (total 182), and recruiting more until each condition had 200 participants (total 4,800). We made the HIT available to U.S. workers with an approval rating of 97\% or more. The HIT carried a reward of \$0.8, which we calculated to ensure that the majority of workers would receive the U.S. minimum wage according to pilot study completion times.

%% file: 5_results.tex
\section{Results}

\subsection{Data Preliminaries}
The average completion task time was 3.6 min (SD: 6.6). 
To analyze participants' responses, we fit the elicited beliefs to a Beta distribution. We treat the elicited point estimate as the mode of a Beta distribution ($\omega$) and the width of the interval as the concentration parameter ($\kappa$) to fit a distribution using optimization as suggested by prior work~\cite{wu2008}. To compute each participant's normative posterior distribution, we used the relationship between the posterior Beta parameters and those of the prior and likelihood deriving from Bayes' rule (Eq.~\ref{eq:simple_bayesian}).

\subsection{Outcome Measures}
We treat the deviation between the participant' actual posterior beliefs and the normative posterior beliefs as a proxy for \textit{how well} the participant appears to have interpreted the information contained in the observed data and combined them with their knowledge they already had.
We analyzed the deviation in two ways. First, to provide intuition for how participants updated in terms of the familiar notions of a distribution's location and variance, we compared the \textit{location} (i.e., mean) and the \textit{variance} of each participants' posterior distribution to those of the normative posterior distribution.

Second, we pre-registered an analysis using KL Divergence (KLD) to measure the difference between a participant's stated posterior beliefs and the normative posterior distribution from our Bayesian models. KLD captures the information loss when representing a target distribution \textit{p} with a second distribution \textit{q}~\cite{kullback1951}.

\subsection{Overview of Updating by Location vs. Variance} \label{sec:updating_loc_var}
We analyzed qualitative differences in how participants updated their beliefs across datasets and visualization conditions. 

\begin{figure*}[htb]
 \centering
 \includegraphics[width=0.80\linewidth]{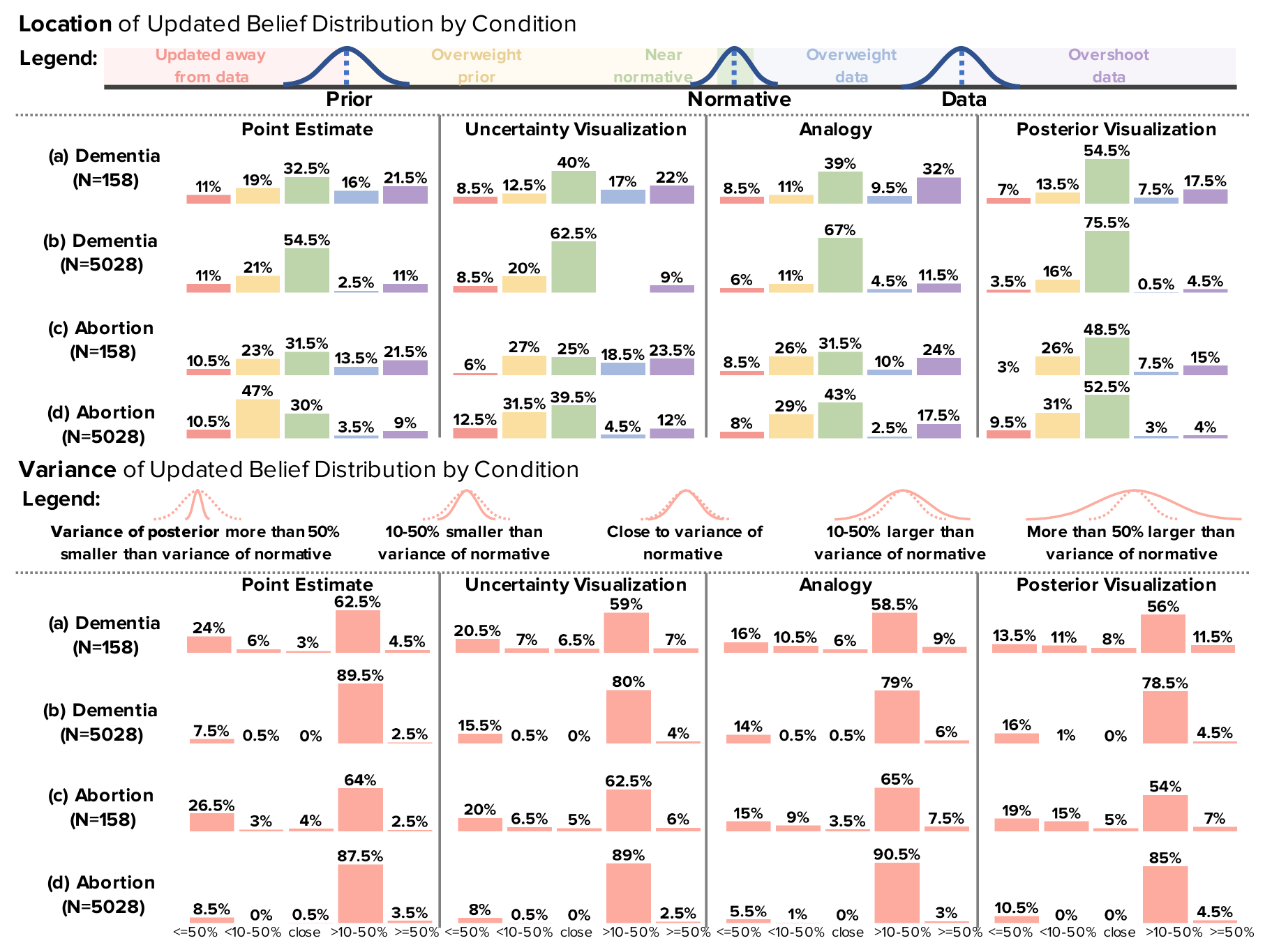}
 \vspace{-5mm}
  \caption{
  Categorization of the location and variance of participants' updates relative to the predictions of normative Bayesian inference for that participant. Top: Each participant was categorized according to the relationship between the mean of their posterior distribution relative to that of their prior distribution, the normative posterior distribution, and the likelihood function. The legend shows a hypothetical participant for which the mean of their prior distribution was smaller than that of the likelihood; our analysis also includes the opposite case (i.e., the mean of the participant's prior was greater than the mean of the likelihood). Bottom: Each participant was categorized according to the relationship between the variance of their posterior distribution relative to that of the normative posterior distribution.}
  \vspace{-4mm}
  \label{fig:update_type}
\end{figure*}

\subsubsection{Location of Updated Belief Distribution by Condition}
We categorize participants into five ``update types'' based on the location (i.e., mean) of their posterior distribution relative to their prior distribution, the normative posterior for that participant, and the likelihood (Fig.~\ref{fig:update_type}). 
We use \textit{near normative} when the location of the participant's posterior is within a relatively small window of the normative posterior (i.e., +/- 2\%). We use \textit{overweight prior} for cases where a participant overweighted their prior distribution relative to the predictions of normative Bayesian updating, and \textit{overweight data} for cases where the participant's posterior fell between the prior and likelihood but was closer to the likelihood than predicted by normative Bayesian updating. While most participants' posterior distributions fell, as we might expect, somewhere between their prior distribution and the likelihood, we use \textit{updated away from data} for cases where participant's posterior moved in an opposite direction from the likelihood as well as their prior. We use \textit{overshoot data} for cases where the location of the participant's posterior surpassed or ``overshot'' the observed data.

Figure~\ref{fig:update_type} characterizes participants' updating behavior by dataset and visualization condition according to these categories. Overall, 
the \textit{near normative} type was the most frequent across datasets and conditions, suggesting that people are approximating Bayesian updating in terms of the location of their distributions.
Participants in the Point Estimate conditions (first column in Fig.~\ref{fig:update_type}) were the least likely to fall in the \textit{near normative} category, and those in the Posterior Visualization conditions (last column) were the most likely to.

Overweighting one's prior was, however, more common in two conditions: the Point Estimate for the large abortion dataset and Uncertainty Visualization for the small abortion dataset. The greater tendency among participants to perceive the abortion dataset as having been manipulated may have led participants to adhere more strongly to their prior beliefs. 

Similarly, when comparing the ratio of the \textit{overweight prior} type between dementia datasets (row a and b) and abortion dataset (row c and d), more participants overweighted their priors when they examined abortion datasets. 

Figure~\ref{fig:update_type} also indicates that the analogy conditions resulted in the highest ratio of people who overshot the likelihood across datasets. The vast majority (roughly 95\%) of our participants had more uncertain priors compared to the likelihood, leading to multipliers greater than one. It is possible that imprecise mental calculations led analogy participants to overcorrect.

\subsubsection{Variance in Updated Beliefs by Condition}
To contextualize how the amount of uncertainty implied by participants' posterior beliefs compared to the amount predicted by normative inference, we categorized patterns in variance updates (Fig.~\ref{fig:update_type}). Because the deviation in elicited posterior versus normative posterior variance was considerably larger than that for means, we categorized participants as \textit{close to normative} if the participant's posterior was within 10\% of the variance of the normative posterior. We similarly categorized participants whose posterior variance was more than 50\% smaller than the variance of the normative posterior, as well as 10-50\% smaller, 10-50\% larger, or more than 50\% larger.

Comparing the distribution across categories in Figure~\ref{fig:update_type} Location (top) to that in Figure~\ref{fig:update_type} Variance (bottom), it is clear that participants' deviations from normative inference are driven primarily by non-Bayesian updating of the variance of their beliefs.
Additionally, in contrast to the results on location updating, we see no clear advantages of the two types of Bayesian assistance in reducing errors in variance updating. Regardless of the specific dataset, most participants provided posterior beliefs the variance of which was 10\%-50\% higher than the variance of the normative posterior. Hence, participants remained more uncertain about the parameter value than they should have in general. Possible drivers of this pattern include unmodeled predictors (e.g., a person's relative trust in data relative to a Bayesian), error in elicitation, or non-Bayesian updating.

Variance results are somewhat different between the small (row a and c) and large datasets (row b and d). Specifically, around 30\% of participants who saw small datasets were more certain than the normative posterior (summing up the first two bars). However, for those who saw large datasets, this number dropped to less than 17\% of participants. Overall, participants were less certain of their updated beliefs than the normative posterior, but those who saw the small datasets were overconfident more frequently than those who saw the large datasets.


\begin{figure}[!t]
 \centering
  \includegraphics[width=0.95\columnwidth]{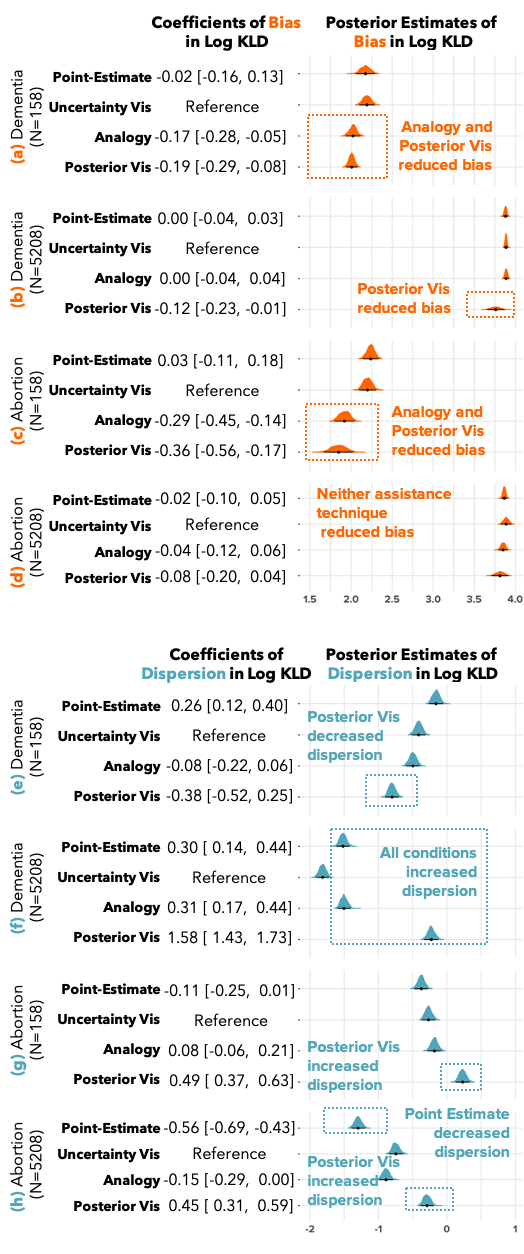}
  \vspace{-6mm}
  \caption{Posterior estimates of bias (mean error) and dispersion (standard deviation) of log KLD with 95\% credible interval by condition.  Results for the dementia datasets are presented in the top row, and for the abortion datasets in the bottom row.  Annotations describe effects relative to visualizing uncertainty in observed data (Uncertainty Vis).
  }
  \label{fig:results_all}
  \vspace{-5mm}
\end{figure}

\subsection{Preregistered Models: Updating by Log KLD}
Per our pre-registration, we specified four Bayesian linear regressions, one for each dataset we presented to participants (dementia N=158, dementia N=5208, abortion N=158, abortion N=5208). These regressions estimate differences in the \textit{distributions of KLD}, a singular measure of deviation between each participant's updating and normative Bayesian updating, by condition.  
\vspace{-2mm}
\begin{eqnarray}
   kld &\sim& \textrm{dlnorm}(\mu, \sigma) \nonumber\\
    \textcolor{bias}\mu &=& \mu_{int} + \mu_{post} * Post  \nonumber\\ 
    &+& \mu_{anlg} * Analogy + \mu_{pointEst} * PointEst  \nonumber\\
    \textcolor{disp}\log(\textcolor{disp}\sigma) &=& \sigma_{int} + \sigma_{post} * Post  \nonumber \\ 
    &+& \sigma_{anlg} * Analogy + \sigma_{pointEst}* PointEst  \nonumber \\ 
    \mu_{int},\mu_{post}, \mu_{anlg}, \mu_{pointEst} &\sim& dnorm(0,5),  \nonumber\\
    \sigma_{int}, \sigma_{post}, \sigma_{anlg}, \sigma_{pointEst} &\sim& dnorm(0,2.5) \nonumber
\end{eqnarray}

Each model consisted of two submodels. The first submodel predicted \bias{bias} (mean error) in log KLD, capturing how closely participants' response distributions aligned with the normative Bayesian prediction by condition.
We use log KLD in our analysis (reporting non-log error results in Supplemental Material) to reduce the impacts of outliers we observed across conditions on our estimates, as KLD grows rapidly as the two distributions diverge more.

The second submodel regressed \disp{dispersion} (variance) in log KLD in log space on the same variables, capturing how much variation there was between participants' deviations from normative inference in a condition. 
In addition to lower bias, lower dispersion (i.e., more consistent) estimates of log KLD means a technique reduces noise.

We implemented each model in R's rethinking package~\cite{rethinking}, using weakly-informed Gaussian prior distributions centered around 0 for bias and dispersion.
We used dummy variables to indicate whether the participant was shown an uncertainty visualization, an analogy, or a posterior visualization. 

We report the result for each condition and dataset relative to a participant in the Uncertainty Visualization condition, as visualizing uncertainty is arguably the best choice a designer could make outside of personalization. We provide  coefficients for both submodels in Figure~\ref{fig:results_all}, left. For readers familiar with statistical significance, we say that a condition has a reliable effect over uncertainty visualization when its 95\% Percentile Interval (PI) (reported in text) does not overlap with 0 (which would indicate the possibility of no effect). We visualize posterior estimates of expected bias and dispersion in log KLD by condition (Fig.~\ref{fig:results_all}, right). Model specifications are in Supplemental Material.

To further contextualize the size of the effects in bias and dispersion, we also report Cohen's d~\cite{cohen2013} and Common Language Effect Size (CLES~\cite{mcgraw1992}), measures of standardized effect size, using our model results. Cohen's d captures the number of standard deviations by which two means differ, while CLES describes what percentage of the time a randomly drawn sample from one distribution would have a higher value than a randomly drawn sample from the second distribution. To calculate effect size on our model estimates, we first constructed an aggregated posterior distribution for each condition, using the bias posterior estimates from the bias submodel and dispersion posterior estimates from the dispersion model. We compute effect size by comparing the distribution of the assistance conditions with that of the Uncertainty Visualization condition.

\subsubsection{Dementia Dataset}
\textbf{Small sample (N=158):} Relative to the Uncertainty Visualization condition, both Bayesian assistance techniques reliably decreased \bias{bias} in log KLD by similar amounts (-0.19, -0.17 respectively; Fig~\ref{fig:results_all}\bias{a}). Viewing a Point Estimate was not distinguishable in log KLD compared to viewing an Uncertainty Visualization. 

Our characterization of updating by location and variance (Sec.~\ref{sec:updating_loc_var}) suggested that the Posterior Visualization helped participants correctly update the location of their beliefs. Hence, the \bias{bias} reduction in log KLD may be driven by better location updating among Posterior Visualization participants. On the other hand, our earlier analysis (Fig.~\ref{fig:update_type}) indicates that the location updating of participants in the Analogy condition and the Uncertainty Visualization condition for the small dementia dataset are similar. Hence the reliable improvement in updating we observe for the Analogy condition may be driven more by better variance updates than better location updating. 

Our \disp{dispersion} submodel indicates that the Posterior Visualization led to more consistent values of log KLD among participants compared to Uncertainty Visualization, with an estimated reduction in dispersion of 0.39 (Fig~\ref{fig:results_all}\disp{e}). Seeing an Analogy did not noticeably affect dispersion compared to the Uncertainty Visualization. However, viewing a Point Estimate increased dispersion in log KLD relative to Uncertainty Visualization. 

Cohen's d for the Posterior Visualization was 0.33, equivalent to a CLES of 59\%. Hence, a participant from Posterior Visualization conditions will have lower log KLD than a participant from the Uncertainty Visualization condition 59 out of 100 times when we randomly select a participant from each condition. Cohen's d for the Analogy assistance was 0.27, equivalent to a CLES 57\%.

\textbf{Large sample (N=5208):} 
Relative to the Uncertainty Visualization condition, viewing a Posterior Visualization reliably reduced \bias{bias} in log KLD, but viewing an Analogy or Point Estimate had no observable effect (Fig.~\ref{fig:results_all}\bias{b}).
 
While highly variant, the distribution of \bias{bias} in log KLD for the Posterior Visualization condition does not overlap with the distributions of expected \bias{bias} for the non-Bayesian conditions (Fig.~\ref{fig:results_all}\bias{b} right).
However, the distribution of expected \bias{bias} for the Analogy condition is not distinguishable from the Point Estimate and Uncertainty Visualization conditions. Again, our earlier analysis of location and variance updates (Fig.~\ref{fig:update_type}) suggests that participants in the Posterior Visualization conditions were better at updating the location of their posterior.

All conditions reliably increased \disp{dispersion} in log KLD relative to Uncertainty Visualization (Fig.~\ref{fig:results_all}\disp{f}) 

Cohen's d for the Posterior Visualization was 0.21, equivalent to a CLES of 56\%.

\subsubsection{Abortion Dataset}
\textbf{Small sample (N=158):} 
Similar to the small dementia dataset, the Analogy and Posterior Visualization both reliably reduced \bias{bias} in log KLD relative to the Uncertainty Visualization (Fig.~\ref{fig:results_all}\bias{c}) 
while the Point Estimate condition was not reliably different.

Compared to the small sample dementia dataset, being in the Posterior visualization condition resulted in higher estimated \disp{dispersion} in log KLD (Fig.~\ref{fig:results_all}\disp{g}). 

Cohen's d for the Analogy and Posterior Visualization were 0.35 (CLES 59\%).

\textbf{Large sample (N=5208):}
In contrast to the large dementia dataset, neither the Posterior Visualization nor the Analogy condition reliably reduced \bias{bias} in log KLD for the large abortion dataset (Fig.~\ref{fig:results_all}\bias{d}). 
A Point Estimate also did not reliably differ from Uncertainty Visualization. 
We suspect that any effects of Bayesian assistance were too small to observe in light of the rather large discrepancies we observed between participants' posterior beliefs and the predictions of normative Bayesian inference with regard to variance (Fig.~\ref{fig:update_type}).

We see slightly different patterns compared to the large sample dementia dataset when it comes to effects on \disp{dispersion} in log KLD. Viewing an Analogy slightly decreased \disp{dispersion} in log KLD 
while viewing a Point Estimate had a stronger decreasing effect (Fig.~\ref{fig:results_all}\disp{h}).  

\subsection{Conceptual Replication of Sample Size Effect}
Our results conceptually replicate a difference in how closely the updates of untrained participants resemble Bayesian updating when shown a small versus a large dataset observed in behavioral economics~\cite{ambuehl2018,benjamin2016} and visual data interpretation~\cite{kim2019}.
While participants assigned large datasets appear to update closer to normative Bayesian inference when we look at location of posterior beliefs (e.g., compare row a and b, and row c and d in Fig~\ref{fig:update_type}), the opposite is true when we look at the variance of their posterior beliefs, where deviation from normative Bayesian inference is substantial.
The average bias in log KLD across participants was 0.90 (median:0.93, IQR:0.23, KLD: 11.24) for small datasets, and much higher for large datasets (mean: 1.67, median:1.68, IQR:0.04, KLD: 49.7), similar to Kim et al.'s~\cite{kim2019} observations for a small sample (n=158) and much larger (n=750k) sample. 

Conceptual models of bias like belief in the law of small numbers~\cite{tversky1971} attempt to explain diverse experimental evidence on belief updating. Our results and those of Kim et al.~\cite{kim2019} are congruent with a model of non-belief in the law of large numbers~\cite{benjamin2016} suggesting that while a Bayesian expects a estimate to eventually converge to the true rate, people update their beliefs as though they expect error in the estimate to be relatively high and constant as sample size increases.


\begin{figure}[htb]
 \centering
  \includegraphics[width=\columnwidth]{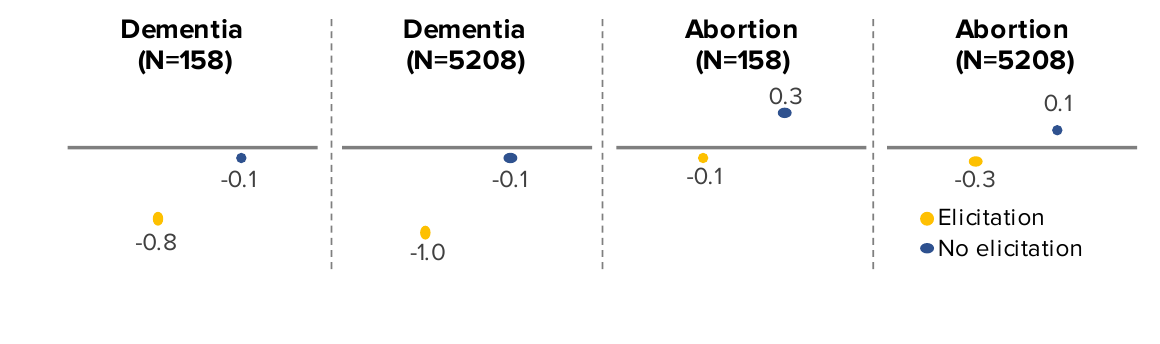}
  \vspace{-0.5in}
    \caption{Comparing whether users from whom we elicited priors updated closer to Bayesian in aggregate than those who did not provide priors. Elicitation conditions yielded lower log KLD, implying prior elicitation alone may improve updating.}
  \label{fig:results_elicitation_effect}
  \vspace{-0.20in}
\end{figure}

\subsection{Effect of Prior Elicitation}
Our results show that conditional on a user specifying their prior, Posterior Visualization and sometimes Uncertainty Analogy better promote Bayesian updating than simply visualizing uncertainty in the observed data. However, given that the status quo in most interactive visualization is not to elicit a prior, one might ask how the act of prior elicitation \textit{itself} impacts updating. Do users become more sensitive to uncertainty in observed data when they explicitly consider their subjective uncertainty about a parameter value? 

Comparing an individual's posterior beliefs to a normative Bayesian posterior with and without elicitation is not possible, as without a prior we would have no way of computing the normative posterior. We instead use an aggregate analysis approach similar to that used in prior work on Bayesian cognition~\cite{griffiths2006,kim2019} and to our approach to computing effect size using CLES (full details reported in Supplemental Material). Across the board, elicitation conditions yielded lower log KLD, suggesting prior elicitation alone may improve updating (Fig.~\ref{fig:results_elicitation_effect}).

%% file: 6_discussion.tex
\section{Bayesian Cognition as Visualization Framework}
We reflect on the potential for using Bayesian assistance and Bayesian modeling to improve visualization.

\subsection{Bayesian Assistance as Design Strategy}
Our work adds to growing evidence that a Bayesian cognition approach can deepen insight into belief formation from visualization and give rise to new design and evaluation techniques for visualization research and practice. 

Our results first provide evidence of tendencies in how untrained users form beliefs from data. Comparing our analysis of location updates to that of variance updates as a whole (Sec.~\ref{sec:updating_loc_var}), it is clear that people are much better at providing posterior beliefs that are located (i.e., have a mean that is) approximately near the location of the normative Bayesian posterior beliefs than they are at providing posterior beliefs that are appropriately certain. Specifically, study participants remained considerably less certain that the information-pooling Bayesian would do, aligning with recent empirically-based models of belief updating from behavioral economics~\cite{benjamin2016} as well as the large sample results of Kim et al.~\cite{kim2019}.


When visualizations present estimates based on small samples for inference, generating Bayesian assistance from users' priors in the context of a simple Bayesian model can improve untrained users sensitivity to how informative new data are. Compared to visualizing uncertainty in an estimate, Bayesian assistance resulted in a small to moderate reduction in bias in updating for estimates based on small samples, even when data were perceived as moderately likely to have been manipulated. When the Bayesian assistance techniques were compared to point estimates, which remain the default approach to presenting estimates in many venues~\cite{hullman2019}, the Bayesian assistance techniques were slightly more effective (CLES from 55\% to 61\%). Using prior beliefs as an entry point into communicating uncertainty via Bayesian assistance may therefore be helpful in common small sample scenarios like presentations of poll results, where people's misinterpretations of uncertainty in data often have implications for their decisions. It can also reduce heterogeneity in updating behavior, especially if the alternative presentation is a point estimate with sample size. 

The benefits of Bayesian assistance for large sample scenarios are less clear-cut. For the dementia dataset, visualizing a predicted Bayesian posterior better aligned participants' posterior beliefs on average with Bayesian inference. This effect, similar to the effects of posterior visualization that we observed for small samples, appears to be driven mostly by the Bayesian assistance helping people more accurately update the location of their beliefs.
We note, however, that the effect of posterior visualization for the large dementia dataset may be too small to be of practical significance, as in a large data case KLD can be sensitive (e.g., even if two highly concentrated distributions are quite close in location, KLD can yield a high value.


The Analogy condition did \textit{not} reliably improve inference for the large dementia dataset. It is possible that people struggled to use large multipliers to arrive at the normative posterior implied by the analogy, as larger numbers are associated with less precise mental representations and more error in mental calculation~\cite{dehaene2011}. 

For the large abortion dataset, which participants rated as slightly more likely to be subject to manipulation, neither of Bayesian assistance techniques improved inferences. This may be due to participants discounting the informativeness of the data based on their perceptions that it might have been manipulated. We present an analysis in Supplemental Material that provides partial support for this explanation.


\subsection{Prior Elicitation as Beneficial}
The benefits of eliciting data-oriented predictions from visualization users have been demonstrated in prior work by Kim, Hullman, and colleagues~\cite{kim2016,kim2017,hullman2018}. Our work extends these findings using a formal Bayesian evaluative framework. One possible explanation, congruent with the findings that eliciting probabilistic predictions improves uncertainty comprehension of Hullman et al.~\cite{hullman2018}, is that interacting with the prior elicitation interface better prepared participants to reason about uncertainty in the observed data. 
Researchers and authors who want to engage visualization users to think more deeply about estimates should consider eliciting subjective uncertainty as an alternative or complement to visualizing uncertainty in estimates.


\subsection{Using Bayesian Inference as Visualization Framework}
Given the potential utility of Bayesian models of cognition to visualization, as demonstrated by our work and prior work~\cite{kim2019,wu2017}, it is worth considering the importance of assumptions of these models and the design requirements of using such approaches. 

\subsubsection{Are the Assumptions of Bayesian Cognition Valid?}
Using Bayesian models of cognition in visualization assumes that users have prior beliefs, they can articulate them when guided to do so, and that greater alignment between how they update their beliefs and how a Bayesian would is desirable (Sec.~\ref{sec:assume}). A common question might be, can I trust the prior beliefs that a participant provides? We refer the reader to literature in economics and psychology for detailed evidence suggesting that people can provide priors unincentivized, and that elicited or inferred representations of people's prior beliefs has predictive value for their later behavior (Sec.~\ref{sec:related}).

When it comes to applications of Bayesian cognition to visualization design and evaluation, even though it is reasonable to believe that elicited priors are not a perfect representation of a user's prior beliefs, we find evidence that they can still be useful to consider in interaction. Prior elicitation itself may be beneficial for prompting a more uncertainty-aware mindset on the part of a visualization user. Moreover, when multiple belief updates by the same person can be observed, as might be the case in visual analytics scenarios, a Bayesian framework can enable detecting patterns of irrational movement or uncertainty reduction in beliefs even if users are far from the predicted Bayesian posterior, due to noise in eliciting prior beliefs or approximate Bayesian behavior~\cite{augenblick2018}. For example, regardless of the distance between their posterior beliefs and normative Bayesian posterior beliefs, if a person increasingly shifts their beliefs without becoming more certain over time, or becomes much more certain without any shifts in beliefs, it is relatively obvious that their belief formation is not responding appropriately to data. It may be worth exploring how prior elicitation could be avoided while still gaining the benefit of Bayesian models for bias detection in visual analytics settings where its reasonable to infer a prior based on data that the system has observed the analyst examining in the past.

By explicitly suggesting to a user how they should update their beliefs in light of new data, Bayesian assistance poses interesting questions about when Bayesian inference is the most appropriate normative standard. 
For example, under what conditions should a user who is distrustful of a data source be guided to integrate the new information into their prior beliefs? While this question is beyond the scope of our work, we believe that there are a number of cases where valid data is rejected irrationally by users, such as when distrust in the source of a media report (e.g., a Conservative leaning publication) leads a Democrat to reject new information that is in fact trustworthy.

In cases where a simple Bayesian model that assumes a user takes data at ``face value'' seems clearly inappropriate, such as when a data source is well known to not be trustworthy, Bayesian modeling can help visualization researchers arrive at a more precise understanding of influences external to the data.
Factors that shape data reception, like the influence of one's a priori trust in the data source, the interaction between the specific parameter estimate and one's beliefs about the source~\cite{austin1994,robinblom}, the tendency to reject one's beliefs entirely upon realizing one was misinformed, or the tendency for people to diverge from a Bayesian's tendency to form posterior beliefs with less variance than their prior or the likelihood even cases where the prior and likelihood would seem disparate are all fair game for including in more sophisticated Bayesian models in the form of ``hyperpriors'' (distributions over parameters of the priors). We believe such ``pseudo-Bayesian'' models could provide the basis for understanding a large class of cognitive biases that affect judgments from visualizations. 

\subsubsection{Generalization of Bayesian Approach}
How to use Bayesian cognition for understanding or improving belief updating from visualizations may at first seem complicated. We suggest that a natural starting place to apply the approach involves first determining what parameter(s) a visualization supports estimating. 
The parameter(s) should correspond to statistics on the observed data that the author believes are most important to the user and inference task: a population-level proportion (rate), a bivariate relationship (with parameters, e.g., of a slope and intercept), an average. 

A Bayesian model can be specified to estimate the posterior probability of the parameter(s) given a prior distribution and likelihood function assumed to characterize data generation. As our experiment demonstrates, even a simple model may suffice to drive improved inferences. While Bayesian modeling is flexible to varying forms of prior and posterior distributions, model specification is often simplified by looking to a family of distributions associated with a type of parameter and likelihood to identify the conjugate prior (e.g., a Beta distribution for probability, a truncated Gaussian for a positive-valued random variable, a Gamma for a duration, etc.). Textbooks aimed at readers new to Bayesian modeling provide accessible explanations and examples of common model formats~\cite{kruschke2014,mcelreath2016}
The Bayesian model we employed for a Binominal likelihood function to generate Bayesian assistance has just a single parameter. However, the general intuition behind Bayesian assistance applies to other data generating processes like Gaussians, where the mean of the normative posterior is the weighted average between the mean of the prior and the observed data weighted by the amount of information in each distribution. More detail on how to calculate posterior parameters when the likelihood function follows other distributions (e.g., Normal distribution) is in Supplemental Material.


We believe that the potential for Bayesian assistance to be used as a design strategy in visualization analysis and communication settings extends far beyond the demonstration we presented here. For example, while we use an individual's prior from a single belief update to drive the two forms of Bayesian assistance, recent work from economics suggests that how a person updates their beliefs in light of new data is a stable individual trait~\cite{atanasov2020small,augenblick2018,dominitz2011,moebius2011managing}. Personalizing data representations based on an individual's ``update type'' (e.g., tendency to overweight vs. underweight their prior or data) may be beneficial in visual analytics or communication settings.

\section{Conclusion}
We showed how personalizing the presentation of visualized data using Bayesian inference can assist untrained visualizations users in updating their beliefs more like Bayesians. Through a large experiment (N=4,800), we found that presenting a Uncertainty Analogy or Posterior Visualization improved belief updating for proportion estimates compared to typical presentations of uncertainty for small datasets, and, in some cases, for large datasets for which people tend to deviate more from normative inference. By comparing to visualizing uncertainty in the data via a shaded interval, we show that better responsive to new information captured by data may require more sophisticated, theoretically-driven approaches like Bayesian cognition. Further, an aggregate level analysis of updating suggested that prior elicitation alone may improve Bayesian reasoning. Our Bayesian framework can be applied to gain insight into belief formation, better define ``normative'' consumption of data visualizations, and guide interactions with data in a range of contexts.

\section{Acknowledgements}
This work was supported by NSF award \#1930642. Many thanks to Peaks Krafft for comments early on, and Abhraneel Sarma and Alex Kale for their helpful feedback on drafts.